\documentclass[10pt,twocolumn]{IEEEtran}
\usepackage{amsmath,amsthm,amssymb,psfig,epsfig,subfigure,cite}
\usepackage{algorithm,algpseudocode}

\newtheorem{theorem}{\hspace*{1pc}Theorem}
\newtheorem{corollary}{\hspace*{1pc}Corollary}

\newtheorem{definition}{\hspace*{1pc}Definition}

\begin{document}

\setlength{\parindent}{1pc}

\title{Loss Tomography from Tree Topologies to General Topologies}
\author{Weiping~Zhu, Ke Deng \thanks{Weiping Zhu is with University of New South Wales, Australia}}
\date{}
\maketitle

\begin{abstract}
Loss tomography has received considerable attention in recent years
and a number of estimators based on maximum likelihood (ML) or
Bayesian principles have been proposed. Almost all of the estimators
are devoted to
 the tree topology despite the general topology is more common in practice. There has been few likelihood function  devoted to the general
  topology, not to mention the estimator.  To overcome this, two sets of sufficient statistics for the tree and general topologies, respectively, are proposed in this paper. Using the statistics, two likelihood functions, one for a topology, are proposed here and subsequently two likelihood equations for
  the general topology,
 one is link-based and the other is path-based, are obtained.
 In addition, a dependence between subtrees in terms of their estimates is identified for the general
topology and a divide-and-conquer strategy is proposed to deal with the dependence, which divides a
general network into two types of
independent trees. Further,  two algorithms, one for a
type of the independent trees, are proposed to estimate the loss rates of each type.
\end{abstract}


\section{Introduction}

Network characteristics, such as loss rate, delay, available
bandwidth, and their distributions, are critical to various network
operations and also important to network research, e.g modeling.
Because of these, considerable attention has been given to network
measurement, in particular to large networks. However, due to various
reasons, e.g. security, commercial interests and administrative
boundary, some of the characteristics cannot be obtained directly from
a large network. To overcome the limitations, network tomography is
proposed in \cite{YV96}, where the author suggests to use
end-to-end measurement and statistical inference to estimate the
characteristics of a large network. The characteristics that have been estimated in this manner
include link-level loss rates \cite{BDPT02}, delay distribution
\cite{LY03}, \cite{TCN03}, \cite{PDHT02}, \cite{SH03}, \cite{LGN06},
and loss pattern \cite{ADV07}. In this paper, our attention is focused
on the loss rate inference, which can be easily extended to estimate
loss pattern.

In an active approach, a number of
sources attached to a network send probes to the
receivers attached to the other side of the network, where the paths
from the sources to the receivers cover the network of interest. To create the correlation required for estimation,
probes need to be sent in a coordinated manner to the receivers, where a multicast scheme is often used to achieve this. Then, the
arrivals, arrival orders and arrival time of the probes are used by an estimator to infer the characteristics of
interest.
 If the probes sent in an experiment are
far apart and the traffic is statistically stable, the observations obtained at
receivers are considered independent identical distributed
($i.i.d.$) and the likelihood function of the
observations is the product of the individual ones.
To
determine the parameter(s) embedded into a likelihood function,  ML or Bayesian principles
are often used in the determination. No matter which principle is used
in estimation, 
the sufficient statistics from observations are crucial for an estimator. Unfortunately, this issue has
been overlooked in the past that leads some of the estimators proposed so far to use a part of the available information.

Apart from statistics, most of works in loss tomography have been focused on the tree topology although a general topology is more common in practice. The tree topology as named has a single
source attached to the root of a multicast tree to send probes to
the receivers attached to the leaf nodes of the multicast tree, where all of the receivers have a common view of the probing process.  A network of the general topology
 is different from the tree one that is composed of a number of trees and some parts of the trees are intersected with each other. Thus, the receivers, nodes, and links in an intersected subtree
can receive probes from multiple sources and observe a number of probing processes. Because of the difference, an estimator for the general topology, in particular for the intersected areas, must consider the probes sent by different sources. In addition, the estimator must consider the possible
 dependence between  intersected areas and other areas. The differences make the estimators or algorithms developed for the tree
topology incapable of the general topology. This paper aims at overcoming the shortage that will   address the statistics, the correlation and the
dependence raised above. In addition, it provides a number of ML estimators
for the general topology.

\subsection{Contribution and Paper Organization}
As stated, there has been a lack of ML estimators in loss tomography
for the tree topology and  there has been a lack of both analytical
results and estimators for the general topology. To fill the gaps, we in this paper present the results obtained recently that partially
solve the problems, which, comparing to \cite{CDHT99}-\cite{ZG05},
have two-fold contributions to loss tomography, one for a topology.
For the tree topology, there are 2 contributions:

\begin{enumerate}
\item A set of the minimal sufficient statistics is introduced and used to rewrite
the widely used likelihood function into a different format. From the
likelihood function, a set of likelihood equations is derived that
shows the loss rate of a link depends on the loss rates of its
ancestors and descendants, where the polynomial proposed in
\cite{CDHT99} is a special case of the likelihood equations.
\item  The solution space is proved to be concave if a Bernoulli model is
assumed to describe the loss behavior of a link. The finding ensures
that
 the estimates obtained by an iterative procedure, e.g.
EM, are MLEs.
\end{enumerate}

\noindent For the general topology, there are four contributions:

\begin{enumerate}
\item The analytical results obtained from the tree
topology are extended to the general topology, and a direct
expression of the MLE is derived for the loss rate of a link. The
direct expression has a similar structure as that of the tree
topology.
\item   By proving the direct expression leading to an MLE, we also prove
the solution space is concave and the estimate obtained by an
iterative algorithm, such as the EM, the fixed point and MCMC, is an
MLE.
\item Apart from the link-based estimator, a path-based estimator is
proposed that not only generalizes the results presented in
\cite{CDHT99}, but also points out the specialty of the general
topology. The specialty rests on the demand of consistency in
estimation that leads to an order in estimation.
\item A
divide-and-conquer strategy
 is proposed to impose the order raised above that decomposes a general network
into two type of independent trees; one is for those without any intersection with others, the other is for those with intersection(s).
Further, two estimators are proposed, one for each of the above.
\end{enumerate}
Apart from the above, a practical issue that has been overlooked by
previous works is raised in this paper, which is about the number of
probes needed to capture the loss rate of a link. In contrast to
previous works, the focus here is not only on the number of probes
that need to be sent but also on the consistency between the
background traffic and the model used by an estimator. Only if the
model used to develop the estimator is consistent with that of the
background traffic, is the accuracy of the estimator guaranteed. If not,
however,  there is no guarantee that an estimate obtained from more
samples is better than that obtained from less ones. Our simulation
study shows that for some background traffic, 5\% variance can be
easily achieved with a few hundreds of samples. Nevertheless, for some
other traffic,  10\% variance is hardly achieved even with a few
thousands of samples. In this case, increasing the percentage of the
probes in traffic can improve estimation
 accuracy, i.e. sacrificing efficiency for accuracy. Despite this, the
 estimate obtained may not be the true one since the probes
 sent can intervene the background traffic.

The rest of the paper is organized as follows. In Section \ref{section2}, we
present the related works and the notations used in this paper. In
addition, a set of sufficient statistics are introduced in this
section. There are four parts in Section \ref{section3}: a) a likelihood function of the tree topology is created from the set of
sufficient statistics; b) the likelihood equation from the likelihood
function is derived; c)  the
likelihood equation derived in b) and that presented previously are compared and the difference between them are identified; and d) the statistics introduced in Section \ref{section2} are proved
to be the complete minimal sufficient ones. We then extend the results
from the tree topology to the general topology in the rest of the
paper. Section \ref{section4} contains the extended notation and the set of
sufficient statistics. Then, the link-based and the path-based
likelihood equations are presented in Section \ref{section5}. Section \ref{section6} is devoted
to the solution of the likelihood equations proposed in Section \ref{section5}.
Statistical properties of the estimators are discussed in Section \ref{section7}.
Section \ref{section8} presents simulation results.
 The last section is devoted to
concluding remark.

\section{Related Works and Problem Formulation}
\label{section2}
\subsection{Related Works}
Multicast Inference of Network Characters (MINC) is the pioneer of
using multicast probes to create correlated observations, where a
Bernoulli model is used to model the loss behaviors of a link. Using
this model, the authors of \cite{CDHT99} derive an MLE for the pass
rate of a path connecting the source to a node. The MLE is expressed
in a polynomial that is one degree less than the number of descendants
of the node \cite{CDHT99}, \cite{CDMT99}, \cite{CDMT99a}.  To ease the
concern of using numeric method to solve a higher degree polynomial $(
> 5 )$, the authors of \cite{DHPT06} propose an explicit estimator.
Although the estimator has the same asymptotic variance as that of the MLE to first
order, it is not an MLE and there is a noticeable difference between
the estimate obtained by the estimator and an MLE if $n<\infty$. The
two estimators are dedicated to the tree topology. Later, Bu {\it et.
al.} attempted to extend the strategy to the general topology without
success in \cite{BDPT02}.  The authors then resorted on iterative
procedures, i.e. the
 EM, to approximate the
 maximum of the likelihood function. In addition, a minimum variance weighted average (MVWA) method is used for comparison with the EM one. The problem of
 MVWA is the bias of its estimate when the sample size is small \cite{Rao80}. The experimental results presented in \cite{BDPT02}
 confirm
this, where the EM outperforms the MVWA when the sample size is small.
Nevertheless, the estimates obtained by the EM algorithm could not be
verified to be the MLE since there was no proof that the search space
is concave. In addition, iterative procedures, e.g. EM algorithm, have
their interionic weakness as previously stated in \cite{DLR77}.
Considering the unavailability of multicast in some networks, Harfoush
{\it et al.} and Coates {\it et al.} independently proposed the use of
the unicast-based multicast to send probes to receivers \cite{HBB00},
\cite{CN00}, where Coates {\it et al.} also suggested the use of the
EM algorithm to estimate link-level loss rates. Apart from those,
Rabbat {\it et al.} in \cite{RNC04} consider network tomography on
general networks and found a general network comprised of multiple
sources and multiple receivers can be decomposed into a number of 2 by
2 components. The authors further proposed the use of the generalized
likelihood ratio test to identify network topology. To improve the
scalability of an estimator, Zhu and Geng propose a bottom up
estimator  for the tree topology in \cite{ZG04}, which later is
found to be topology independent \cite{ZG05}. The estimator adopts a
step by step approach to estimate the loss rate of a link, one at a
time and from bottom up. At each step the estimator uses a formula to
compute the loss rate of a link. Despite the effectiveness,
scalability, and extensibility to the general topology,
 the estimate obtained by the estimator is not the MLE as the one proposed in \cite{DHPT06} because the statistics used
 by the two estimators are not sufficient ones.

\subsection{Notation}

\begin{figure}
\centerline{\psfig{figure=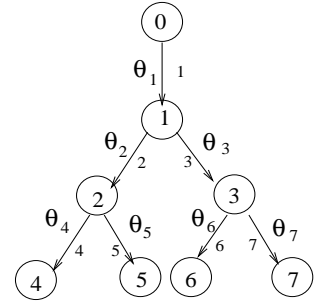,height=4.0cm,width=4cm}}
\caption{Network structure } \label{struct}
\end{figure}

 Let $T=(V, E, \theta)$ donate the multicast tree, where
$V=\{v_0, v_1, ... v_m\}$ is the set of nodes representing routers and
switches of a network; $E=\{e_1,..., e_m\}$ is the set of directed
links connecting the nodes of $V$, the two nodes connected by a link
are called the parent node and the child node of the link, where the
parent forwards probes received from its parent to the child; and
$\theta=\{\theta_1,..., \theta_m\}$ is the set of parameters to be
estimated, one for a link. The multicast tree used to deliver probes
to receivers is slightly different from an ordinary one at its root
that has only a single descendant. Figure \ref{struct} presents an
example of the multicast tree.

To distinguish a link from another, each link is assigned a unique
number from 1 to m; similarly, each node also has a unique number from
0 to m, where link $i$ is used to connect node $i$'s parent node to
node $i$. The numbers assigned to the nodes are from small to big
along the path of the Breadth-first traversal. The source attached to
node 0 sends probes to the receivers attached to the leaf nodes. $R$
is used to denote the receivers attached to T. In contrast to
\cite{CDHT99} and \cite{DHPT06} that use node as reference in
discussions, we use link instead because there is no one-to-one
correspondence between nodes and links in the general topology.
 As a hierarchical structure, each
link in a tree except the root and leaves has a parent and a number of
descendants. If $f_1(i)$, simply $f(i)$ later, is used to denote the
parent link of link $i$ and $f_l(i)$ to denote the ancestor that is
$l$ hops away from link $i$ in the path to the root, we have
$f_k(i)=f(f_{k-1}(i))$. In addition, let $a(i)=\{f(i),
f_2(i),\cdot\cdot, f_k(i)\}$, where $f_{k}(i)=1$, denote the ancestors
of link $i$. Further, $d_i$ denotes the descendants of link $i$ and
$|d_i|$ denotes the number of descendants in $d_i$. To distinguish
subtrees, each multicast subtree is named by the number assigned to
its root link, where $T(i)=\{V(i), E(i), \theta(i)\}, i \in
\{1,\cdot\cdot, m\}$ denotes the multicast subtree rooted at node
$f(i)$, where $V(i)$, $E(i)$ and $\theta(i)$ are the nodes, links and
parameters of $T(i)$. Note that $T(i), i \notin R$,  is rooted at node
$f(i)$ that uses link $i$ to connects subtree $i$. The group of
receivers attached to $T(i)$ is denoted by $R(i)$. If $n$ probes are
dispatched from the source, each probe $i=1,...., n$ gives rise of an
independent realization $X^{i}$ of the loss process $X$. Let $X_k^i$
denote the state of link $k$ for probe $i$, $X_k^i=1, k\in E$ if probe
$i$ passes link $k$; otherwise $X_k^i=0$.
$\Omega=(X_k^{i})^{i=1,2,...,n}, k \in R$ comprise the observations of
an experiment. In addition, $\Omega_j$ is the observations of $R(j)$
and $(Y_k^{i})^{i=1,2,...,n}, k \in E$ is the state of link $k$
inferred from observation $i$. $Y_k^i=1$ if at least one of $R(k)$ receives probe $i$, otherwise $Y_k^i=0$.

\subsection{Internal State and Internal View}
Given the loss model of a link,  statistical inference is used to
determine the unknown parameters of the likelihood function. If the ML
principle is applied, the task is frequently described as the
maximization of the following log-likelihood function \cite{CDHT99}:
\begin{eqnarray}\label{obj}
\arg\max_{\theta\in\Theta}L(\theta)= \arg\max_{\theta\in\Theta}\sum_{x
\in \Omega}n(x)\log P(x; \theta)
\end{eqnarray}

\noindent where $\Theta=[0,1)^m$ is the value space of parameter
$\theta$, $n(x)$ is the number of observation $x, x \in \Omega$, and
$P(x;\theta)$ is the joint probability of observation $x$ with
parameter $\theta$. However, it is hard, if not impossible, to derive the likelihood equation from a log-likelihood function in the form of
(\ref{obj}).

\label{internal view1}
 Instead of using the log-likelihood function as
(\ref{obj}), we consider to rewrite it in a different form. Let
$P_\Omega(\theta)$ denote the likelihood function. Under the {i.i.d.}
assumption of probes, we have
\[
P_{\Omega}( \theta)=\prod_{i=1}^n P(X^{(i)}, \theta) \]
 and under the Bernoulli assumption of losses at a link, we have
 \[
 P(X^{i}, \theta)=\prod_{j \in E1} (1-\theta_j)\prod_{k \in
 E2}(\theta_k+(1-\theta_k)(1-\beta_k))
 \]
for $X^{i}$, where $E1$ denotes the set of links that have $Y^i_j=1$; $E2$
denotes the set of links whose observation of the probe cannot be confirmed
but their parents' are confirmed, i.e. $Y^i_{f(k)}=1$ and $Y^i_k=0$; and $\beta_i=P(\bigvee_{r\in R(i)} X_r =1|X_i=1;\theta)$ is the pass
rate of subtree $i$. If we do not distinguish the difference between
$E1$ and $E2$ and let $P_{ij}(\theta_j)$ denote the probability of
link $j$ for $X^{i}$, we have
\[
P_{\Omega}( \theta)=\prod_{i=1}^n \prod_{j \in E} P_{ij}(\theta_j)
\]
With a finite $n$ and $|E|$, the order of the products is commutable. To swap the order of the two products, we have
\begin{equation}
P_{\Omega}( \theta)=\prod_{j \in E}\prod_{i=1}^n  P_{ij}(\theta_j).
\label{newlikelihood}
\end{equation}
Having a likelihood function as above, we need to have the state
of each link for each probe, which results in two concepts called
internal state and internal view.  The formal definitions of them are
as follows.
\subsubsection{Internal State}

 Given $X^j$, one is
able to obtain $X_k^{j}, k \in R(i)$ and then we can have
\[
Y_i^j=\max_{k \in R(i)} x_k^j, \mbox{\hspace{1cm}}  j \in \{1, ..,
n\}\] \noindent called the internal state of link $i$ for probe $j$.
If $Y_i^j=1$, probe $j$ for sure passes link $i$.  Further,
considering the values of $Y_i^j$ and $Y_{f(i)}^j$, $i, f(i) \in E$
together, we have three possible combinations:
\begin{enumerate}
\item  $Y_i^j=Y_{f(i)}^j=1$, this also means that probe $j$ passes link
$i$; or
\item  $Y_i^j=0$ and $Y_{f(i)}^j=1$, this means that probe $j$ reaches node  $f(i)$ and then become uncertain in
$T(i)$, i.e. it is not sure whether the probe is lost on link $i$ or
lost in the subtrees rooted at node $i$; or
\item $Y_i^j=Y_{f(i)}^j=0$, this means the probe becomes uncertain at one of $a(i)$ and the uncertainty is transferred from the ancestor to
node $i$.
\end{enumerate}

\noindent For link $i$, if the first occurs, we need to have
$(1-\theta_i)$ in the likelihood function for this probe; if the
second one occurs, we need to have $(\theta_i +
(1-\theta_i)(1-\beta_i))$ in the likelihood function; and if the last
one occurs, we need to have 1 in the likelihood function for this
probe since $P(Y_i^j=0|Y_{f(i)}^j=0)=1$. Because the likelihood
function is in a product form,  we only need to consider the first two in
the likelihood function.
\subsubsection{Internal View}

 Accumulating the states of each link in an
experiment, and let
\[ n_i(1)=\sum_{j=1}^n Y_i^j, \] \noindent  we have the number
of probes passing link $i$ confirmed from $\Omega$, i.e. at least
$n_i(1)$ probes pass link $i$ in the experiment. In addition, let
\[
n_i(0)=n_{f(i)}(1)-n_i(1)
\]
\noindent denote the number of probes that become uncertain in $T(i)$.
$n_i(1)$ and $n_i(0)$ are called the internal view of link $i$. We
will prove that $n_i(1), i\in E $ is a set of sufficient statistics
later in the paper. Using $n_i(1), i\in E $, we are able to have a
likelihood function of $\Omega$ in the format of
(\ref{newlikelihood}).

\section{New Likelihood Function and Solution}
\label{section3}
\subsection{New Likelihood Function} \label{2.a}

Given $n_i(1)$ and $n_i(0), i \in E$, we then have the following
theorem to write the log-likelihood function.

\begin{theorem}\label{sufficient statistics}
Given $n_i(1), n_i(0)$, the log-likelihood function of an
experiment on the tree topology can be written as follows:
\begin{eqnarray}
L(P_{\Omega}(\theta)) &=&\sum_{i\in E}\big[n_i(1)\log(1-\theta_i)+ \nonumber \\
&& n_i(0)\log(\theta_i+(1-\theta_i)(1-\beta_i))\big] \label{newml}
\end{eqnarray}
where  $\theta_i$ and $(1-\beta_i), i \in
\{1,\cdot\cdot,m\}$ are the loss rate of link $i$ and the loss rate of subtree $i$, respectively.
\begin{IEEEproof}
Based on the definition of $n_i(1)$ and $n_i(0)$, we have
$$P(X^j;\theta)=\prod_{i\in E}\big[(1-\theta_i)^{Y_i^j}(\theta_i+(1-\theta_i)(1-\beta_i))^{(Y_{f(i)}^j-Y_i^j)}\big].$$
Thus,
\begin{eqnarray}
L(P_{\Omega}(\theta))&=&\sum_{k=1}^n\log P(X^k;\theta)\nonumber\\
&=&\sum_{k=1}^n\sum_{i\in E}\big[Y_i^j \log(1-\theta_i)+ \nonumber \\
&& (Y_{f(i)}^j-Y_i^j)\log(\theta_i+(1-\theta_i)(1-\beta_i))\big]
\nonumber
\\
&=& \sum_{i\in E}\big[n_i(1)\log(1-\theta_i)+ \nonumber \\ &&
n_i(0)\log(\theta_i+(1-\theta_i)(1-\beta_i))\big]. \nonumber
\end{eqnarray}
The theorem follows.
\end{IEEEproof}
\end{theorem}

If the derivative of (\ref{newml}) can be obtained and expressed
explicitly, the likelihood equation of the tree topology becomes
available.

\subsection{Likelihood Equations and Solution}\label{tree likelihood}

Differentiating (\ref{newml}) with respect to (wrt.) each parameter
and letting the derivatives be 0, we have a set of likelihood
equations as:

\begin{eqnarray}
&&\frac{\partial
L(P_{\Omega}(\Theta))}{\partial\theta_i}=-\frac{n_i(1)}{1-\theta_i}+\frac{n_i(0)\beta_i}{\theta_i+(1-\theta_i)(1-\beta_i)}+
\nonumber
\\&& \beta_i\sum_{k \in a(i)} \frac{n_k(0)\prod_{\substack{l \in a(i) \\ l \geq k}}[(1-\theta_l)\dfrac{1-\beta_{f(l)}}{\theta_l+(1-\theta_l)(1-\beta_l)}]}{\theta_k +(1-\theta_k)(1-\beta_k)} \nonumber \\
&&=0, \mbox{\hspace{2.5cm}} i=1, \cdot\cdot, m. \nonumber
\end{eqnarray}

\noindent  Reorganizing it, we have

\begin{equation} \label{general}
\theta_i=\left\{
  \begin{array}{ll}
  1-\dfrac{\dfrac{n_i(1)}{n}}{\beta_i}, & i=1 \\
 1- \dfrac{\dfrac{n_i(1)}{n_{f(i)}(1)+imp(f(i))}}{\beta_i}, & i \in E\setminus(L\cup 1) \\
 \dfrac{ n_i(0)+
imp(f(i))}{n_{f(i)}(1)+imp(f(i))}, & i \in L
\end{array} \right.
\end{equation}

\noindent where $L$ denotes the set of links connecting $R$. It is
clear that (\ref{general}) can be solved by approximation if $\beta_i$
is a function of $\theta_i$, where

\begin{eqnarray}
imp(f(i))=\sum_{k \in a(i)}\dfrac{n_k(0)*pa_i(k)*\xi_i*\prod_{\substack{l \in a(i) \\
l \geq k}}\dfrac{1-\beta_{f(l)}}{\xi_l}}{\xi_k} \nonumber \\
\label{ancterm}
\end{eqnarray}

\noindent is the estimated number of $n_j(0), j \in a(i)$ that reaches
node $f(i)$ before being lost in $T(i)$, where
\begin{eqnarray}
 \xi_i&=&\theta_i+(1-\theta_i)(1-\beta_i) \nonumber \\
 \beta_i&=&1-\prod_{j \in d_i} \xi_j \nonumber \\
 pa_i(k)&=&\prod_{\substack{l \in a(i) \\
l \geq k}} (1-\theta_l).\nonumber \\
\end{eqnarray}

\noindent  Each term in the summation of (\ref{ancterm}) is for an
ancestor $j, j \in a(i)$ and represents the portion of $n_j(0)$ that reaches link $i$, where $pa_i(k)$ is the pass rate of the path from node
$f_k(i)$ to node $f(i)$; and $\dfrac{n_k(0)}{\xi_k}$ is the estimate
of the number of probes reaching node $f_k(i)$.

\subsection{Similarity and Difference from Previous}

Let $\beta_i=1$, $\forall i, i\in L$, the three formulae of
(\ref{general}) become one as:
\begin{equation}
\theta_i= 1-\dfrac{\gamma(a(i))}{\beta_i}, \label{3to1}
\end{equation}

\noindent where $\gamma(a(i))$ is the pass rate of $T(i)$, its
empirical pass rate is equal to
$\hat\gamma(a(i))=\dfrac{n_i(1)}{n_{f(i)}(1)+imp(f(i))}$. As stated,
if $\beta_i$ is a function of $\theta_i$, $\theta_i$ can be obtained from {\ref{general}}.
In fact, the expression of $\beta_i$ depends on the observation of
$R(i)$. If the observation satisfies
\begin{equation} \label{difference}
\{ (\sum_{l=1}^n \bigwedge_{k \in j} Y_k^l) > 0 |\forall j,  j \in 2^{d_i}\setminus \{\emptyset,\{e\}\} \},
\end{equation} where $2^{d_i}$ stands for the set of all subsets of $d_i$ and $\{e\}$ denotes all single element set, we have,
\begin{equation}
\beta_i=1-\prod_{j \in d_i}(1-\frac{\hat\gamma(a(j))}{1-\theta_i}).
\label{polyequal}
\end{equation}

\noindent If knowing $\hat\gamma(a(i))$, we have
$\hat\beta_i=\dfrac{\hat\gamma(a(i))}{1-\theta_i}$, and then
(\ref{polyequal}) turns to a polynomial as that derived in
\cite{CDHT99}, i.e.

\begin{equation}
H_k(A_k, \gamma)=1-\dfrac{\gamma_k}{A_k}-\prod_{j \in d_k}
(1-\dfrac{\gamma_j}{A_k})=0 \label{minc}
\end{equation}

\noindent Note that (\ref{minc}) is not equivalent to (\ref{general}) since (\ref{general}) is not restricted to (\ref{difference}). If (\ref{difference}) does not hold, $\beta_i$ can not be expressed by (\ref{polyequal}). For instance, if $Y_k, k \in d_i$ has no intersection with others, we need to have a different $\beta_i$ rather than (\ref{polyequal}) in (\ref{general}). There are various $\beta_i$s, depending on observations. For those who are interested in the detail, please refer to \cite{Zhu11}. 

\subsection{Proof of Sufficient Statistics}
\label{mlesection}

There are a number of approaches to prove (\ref{general}) is the
MLE. One of them is to prove the equivalence between (\ref{general})
and (\ref{minc}) that has been achieved in the previous section. An alternative is to
prove the internal view proposed in this paper is the complete minimal
sufficient statistics, To prove this, we need to prove that (\ref{newml}) belongs
to the exponential families. That is obvious since (3) follows a Bernoulli distribution. Alternatively, according to
 the following definition presented in \cite{HC95},

\begin{definition}\label{complete minimal sufficient statistics}
Let $x=\{X_1,....,X_n\}$ be a random sample, governed by the
probability mass function $f(x|A_k)$. The statistic $T(x)$ is
sufficient for $A_k$ if the conditional distribution of $x$, given
$T(x)=t$, is independent of $A_k$.
\end{definition}
we have theorem \ref{sufficient theorem} that confirms the internal views are sufficient statistics.
\begin{theorem} \label{sufficient theorem}
$n_i(1), i \in V$ is a set of sufficient statistics.
\end{theorem}
\begin{IEEEproof}
Given the loss process yields the Bernoulli model and $T(x)=n_k(1)$,
(\ref{newml}) shows the likelihood function is as follows:
\[
f_{A_k}(x)=A_k^{n_k(1)}(1-A_k\beta_k)^{n-n_k(1)}.
\]
The distribution of $n_k(1)$ in $n$ independent Bernoulli trials is a
binomial as
\[
f_{A_k}(x)=\binom{n}{n_k(1)}A_k^{n_k(1)}(1-A_k\beta_k)^{n-n_k(1)}.
\]
Then, the conditional distribution can be written as
\begin{eqnarray}
f_{A_k|n_k(1)}(x)&=&\dfrac{A_k^{n_k(1)}(1-A_k\beta_k)^{n-n_k(1)}}{\binom{n}{n_k(1)}A_k^{n_k(1)}(1-A_k\beta_k)^{n-n_k(1)}.
} \nonumber \\
=\dfrac{1}{\binom{n}{n_k(1)}}.
\end{eqnarray}
According to definition \ref{complete minimal sufficient statistics}, $\{n_1(1),\cdots,n_m(1)\}$ are sufficient statistics. Since
$f_{A_k}(x)$ is of the standard exponential family, the statistics are
minimal complete sufficient ones.
\end{IEEEproof}

\section{Loss Rate Analysis for General Networks}
\label{section4}
\subsection{Goals and Background}
 As stated, using
multiple trees to cover a general network makes estimation harder than
that in the tree topology. Because of this, no likelihood equation has yet
been proposed for the general topology. Without a likelihood equation,
it is impossible to know the shape of the search space of an iterative
algorithm, e.g. the EM algorithm. Subsequently, there is no guarantee
that the estimate obtained by an iterative method is the MLE. To
overcome this, two likelihood equations for the general topology are
presented in the next section: one is a link-based estimator and the
other is a path-based one. The former has a similar structure as
(\ref{general}) that clearly shows the independency of the probes sent
by multiple sources on the estimate of a shared link; the latter
defies previous approaches by considering the number of probes
reaching a node from multiple sources.
Subsequently,
 the path-based likelihood equation is obtained and the solution space of the equation is proved to be strictly
concave. Thus, there is a unique solution to the likelihood equation
that is the MLE.

\subsection{Extended Notation}

 Let ${\cal N}$ denote a general
network consisting of $k$ trees, where $S=\{s^1, s^2, \cdot\cdot,
s^k\}$ denote the sources attached to the trees. In addition, $V$ and
$E$ denote the nodes and links of ${\cal N}$. Each node is assigned a
unique number, so is a link. $|V|$ and $|E|$ denote the number of
nodes and the number of link of the network, respectively.  Each of
the multicast trees is named after the number assigned to its root
link, where $T(i), i \in E$ denotes the subtree with link $i$ as its
root link. Further, let $R(s), s \in S$ denote the receivers attached
to the multicast tree rooted at $s$ and let $Rs(i), i \in E$ denotes
the receivers attached to the multicast subtree rooted at link $i$.
The largest intersection between two or more trees is called the {\it
intersection} of the trees and the root of an intersection is called
the {\it joint node} of the intersection. An intersection is an
ordinary tree and named after its root node. Let $J$ denote all joint
nodes and let $S(j), j\in V \setminus S$ denote the sources that send
probes to node $j$. Since each node can have more than one parents in
a general network, $f_1^s(i)$, simply $f^s(i)$ later, is used to
denote the parent of node $i$ on the way to source $s$ and $f_l^s(i)$
to denote the ancestor that is $l$ hops away from node $i$ in the path
to source $s$. Recursively, we have $f_k^s(i)=f^s(f_{k-1}^s(i))$.
Further, let $a(s,i)=\{f^s(i), f_2^s(i),\cdot\cdot, f_k^s(i)\}$, where
$f_{k+1}^s(i)=s$, denote the ancestors of node $i$ in the path to $s$.
Let $a(i)=\{a(s,i), s \in S(i)\}$ denote all the ancestors of node
$i$.

 If $n^s, s \in S$ denotes the
number of probes sent by source $s$, each probe $o=1,...., n^s$ gives
rise of an independent realization $X^s(o)$ of the probe process $X$.
As the tree topology, $X_k^s(o)=1, k\in E$ if probe $o$ sent by source
$s$ passing link $k$; otherwise $X_k^s(o)=0$. Further,
$\Omega=\{\Omega_s, s \in S\}$, $\Omega_s=(X^s(o)), {o=1,2,...,n^s}$,
comprises the data set of estimation. As the tree, let $Y_k^s(i),
{i=1,2,...,n^s}, k \in E, s \in S$ denote the state of link $k$
obtained from $\Omega_s$ for probe $i$. $Y_k^s(i)=1$ if the probe
passes link $k$, otherwise $Y_k^s(i)=0$. Further, let $\Omega_s(i)$
denote the observation obtained by $Rs(i)$ from the probes sent by
source $s$.

\subsection{Sufficient Statistic}
\label{general sufficient stat}
 It is assumed that each source sends
probes independently, and the arrivals of probes at a node or a
receiver are also assumed independent. Then, the loss process of a
link is considered an {\it i.i.d.} process. This also makes the
collective impacts of probes sent by the sources to a link {\it
i.i.d.}. Thus, the likelihood function of an experiment takes either a
product form of the individual likelihood functions or a summation
form of the individual log-likelihood functions.

 Using the internal view proposed in the tree topology, the confirmed passes of link $i$ for the probes sent by source $s$ can be obtained from
 $\Omega_s$ that is equal to

\[ n_i(s, 1)=\sum_{j=1}^{n^s} Y_{i}^s(j). \] \noindent In addition, let
\[ n_i(s, 0)=n_{p(s,i)}(s,1)-n_i(s, 1)\]

\noindent be the number of probes sent by source $s$ that become
uncertain in $T(i)$. Further, considering all sources, we have
$$n_{i}(1)=\sum_{s \in S(i)}n_{i}(s,1)\hspace{1cm}n_{i}(0)=\sum_{s \in S(i)}n_{i}(s, 0)$$
\noindent be the total number of probes confirmed from $\Omega$ that
pass link $i$ and the total number of probes that turn to uncertain in
$T(i)$, respectively. Using the same method as that presented in
section \ref{mlesection}, we can prove $n_i(1), i \in E$ is a set of
sufficient statistics.

\section{ Maximum Likelihood Estimator}
\label{section5}
Given the similarity between the statistics presented in the last
section and that presented in section \ref{internal view1}, a
link-based estimator can be obtained by a similar procedure as that
presented in section \ref{tree likelihood}.
\subsection{Link-based Estimator}

Using the set of sufficient statistics, we can write the
log-likelihood function of $\Omega$, and using the same strategy as
\cite{Zhu06}, we have the log-likelihood function of a general network
as follows:
\begin{eqnarray}\label{logL-mutisource}
L(\theta) &=&\sum_{i\in{E}}\Big[n_{i}(1)\cdot
log(1-\theta_{i})+n_{i}(0)\cdot
         log\xi_{i}\Big] \label{general tree}
\end{eqnarray}
\noindent where $\xi_i = \theta_i+(1-\theta_i)(1-\beta_i))$.
Differentiating $L(\theta)$ {\it wrt.} $\theta_i$ and setting the
derivatives to 0, we have a set of equations as follows:

\begin{eqnarray}
\theta_i=\left\{
  \begin{array}{l}
  1-\dfrac{\dfrac{n_i(S(i), 1)}{n^i}}{\beta_i},   \mbox{\hspace{2cm}       } i\in RL, \\
  1-\dfrac{\dfrac{ n_i(S(i),1)}{n_{f(i)}(S(i),1)+imp(S(i),f(i))}}{\beta_i}, \mbox{} i \in SBRL, \\
  1-\dfrac{\dfrac{\sum_{j \in S(i)} n_i(j,1)}{\sum_{j
\in S(i)}[n_{f(i)}(j, 1)+imp(j, f(i))]}}{\beta_i}, \\   \mbox{\hspace{4cm}} i\in SSNL, \\
\dfrac{\sum_{j \in S(i)}[ n_i(j,0)+ imp(j, f(i))]}{\sum_{j \in
S(i)}[n_{f(i)}(j, 1)+imp(j,f(i))]},  \mbox{\hspace{0.1cm}     } i \in
AOL,
\end{array}\right. \label{mgeneral}
\end{eqnarray}

\noindent where RL denotes root links, SBRL denotes the links that are
not root link but only receive probes from a single source, SSNL
denotes the links in a intersection but leaf ones, and AOL denotes all
others, i.e. leaf links. As (\ref{general}), $imp(j, i)$ denotes the
impact of $n_k(j,0)$, $k \in a(j,i)$,
 on the loss rate of link $i$.
\begin{eqnarray}
&&imp(j,i)= \nonumber \\
&&\sum_{k \in a(j,i)}\dfrac{n_k(j,0)\cdot pa_i(k)\cdot\xi_i*\prod_{\substack{l \in a(i) \\
l \geq k}}\prod_{q \in C_l \setminus l} \xi_q}{\theta_k
+(1-\theta_k)\prod_{q \in C_{k}}\xi_q} \nonumber
\end{eqnarray}
\noindent where
\begin{eqnarray}
&&pa_i(k)=\prod_{\substack{l \in a(j,i) \\
l \geq k}} (1-\theta_l). \nonumber
\end{eqnarray}

In fact, the second equation of (\ref{mgeneral}) can be merged into
the third one since each link in SBRL has its $|S(i)|=1$. To
distinguish SBRL from SSNL here is for later to divide a general
network into a number of independent trees at $j, j \in J$.

The estimate obtained from (\ref{mgeneral}) can be proved to be the
MLE by using the same procedure as that presented in section
\ref{mlesection}. In addition, almost all of the results
obtained in the tree topology can be extended to the general
topology. For instance, let $\gamma_i(a(i))$ be the pass rate of
$T(i)$, the empirical pass rate of $\gamma_i(a(i))$ is
\begin{figure}
\centerline{\psfig{figure=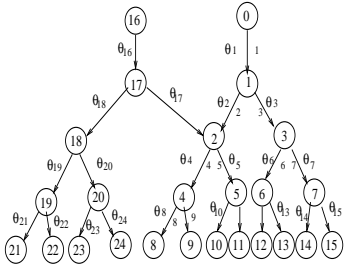,height=3.5cm,width=4.5cm}}
\caption{A Network with Multiple Sources} \label{multisource}
\end{figure}

\begin{equation}
\hat\gamma_i(a(i))= \dfrac{\sum_{j \in S(i)} n_i(j,1)}{\sum_{j \in
S(i)}[n_{f(i)}(j, 1)+imp(j, f(i))]}. \label{mlink-based}
\end{equation}
In addition, if $\beta_i$ can be expressed as a function of
$\theta_i$, the likelihood equation is expressed by a polynomial as
(\ref{polyequal}). However,  there is a lack of methods to express
$\beta_i$ for the general topology because there can have a number of
sources sending probes to  subtree $i$.

\subsection{Insight and Remark}

Despite it is unclear how to express $\beta_i$, the similarity between the four equations presented in (\ref{mgeneral}) and
 the three in
(\ref{general}) provides such a hope that a path-based estimator as
(\ref{minc}) is available to the general topology. To make this
happen, equations (\ref{general}), (\ref{mgeneral}) and (\ref{minc})
are examined and an insight is emerged
 that is presented in the following remark:

{Remark:} regardless of the topology and the number of sources, the
MLE of the loss rate of a link can be obtained if we know:
\begin{enumerate} \item the total number of probes reaching the parent node of
the link, e.g. $n^i$, $n_{f(i)}(S(i), 1)+imp(S(i),f(i))$, or $\sum_{j
\in S(i)}[n_{f(i)}(j,1)+imp(j, f(i))]$ of (\ref{mgeneral});
\item the total number of probes reaching the receivers via
the link, e.g. $n_i(S(i),1)$,  or $\sum_{j \in S(i)} n_i(j,1)$ of
(\ref{mgeneral}); and,
\item the pass rate of the subtree rooted at the child node of the link, e.g. $\beta_i$ of (\ref{mgeneral}).
\end{enumerate}
Note that the link stated in the remark can be a path consisting of a
number of links serially connected. For the tree topology, there is
only one source, 1) and 2) can be suppressed to the pass rate of the
path connecting the source node to an internal node. (\ref{minc}) as
an example takes advantage of this to express $1-\beta_i$, the loss
rate of subtree $i$, in two different  functions of $A_i$: a) the
product of the loss rates of the multicast subtrees rooted at node
$i$; and b) $1-\dfrac{\gamma_i}{A_i}$ since $A_i\beta_i=\gamma_i$.

\subsection{Path-based Estimator}

As stated, there has been a lack of likelihood function proposed for
the general topology regardless of whether it is a link-based or a path-based. This is partially due to the lack of sufficient
statistics of a link or a path from observations. With the help of the
minimal sufficient statistics $n_i(1), i \in V$, a link-based estimator has
been presented in the paper. Further, using the remark presented above and the statistics,  we are able to have
a path-based
log-likelihood function for the general topology. Among the three factors listed in the remark, two of them are
available, i.e.:

\begin{itemize}
\item the total number of probes sent by a source; and
\item the total number of probes reaching $Rs(i)$ and know the source of each arrived
probe.
\end{itemize}
\noindent Given two of the three,  a likelihood equation with the
third factor as the variable can be derived.  Let
\begin{equation}
 A(s,i)=\prod_{j \in a(s,i)}(1-\theta_j), \mbox{\hspace{0.5cm}}  i \in V, \mbox{ and }  s
 \in S(i)
 \label{theta2A}
 \end{equation}

 \noindent be the pass rate of
the path from source $s$ to node $i$. It is easy to prove that
(\ref{theta2A}) is the bijection, $\Gamma$, from $\Theta$ to $A$,
where $\Theta$ is the support space of $\{\theta_i, i \in E\}$ and $A$
is the support space of $\{A(s,i), i \in V\setminus S, s \in S(i)\}$.
To have a polynomial likelihood equation, we need to express $\beta_i$
by $A(s, i)$.

In contrast to the tree topology, there are intersections in the
general topology.  For intersection $i$,  $\beta_i$ must be consistent for all
paths that connect a sources to node $i$. To prove that is a necessary condition for the
MLE, we have the following theorem.

\begin{theorem} \label{general MLE}
The likelihood equation describing the pass rate of a path connecting
source $s$ to node $i$, $i \in J$ can be expressed as
\begin{eqnarray}
A(s,i)=\dfrac{\gamma_i(s)}{\beta_i}, \mbox{\hspace{1cm}} s \in S(i).
\label{pathsetpoly}
\end{eqnarray}
As the tree topology, the empirical probability of $\gamma_i(s)$ can
be obtained by $\dfrac{n_i(s,1)}{n^s}$.
\end{theorem}
\begin{IEEEproof}
Using the sufficient statistics, we can write the path-based
likelihood function as

\begin{eqnarray}
P(A(s,i))&=&\prod_{i \in V\setminus S}\prod_{s\in S(i)} \Big
[A(s,i)^{n_i(s,1)}(1-A(s,i)+ \nonumber
\\
&&A(s,i)(1-\beta_i))^{(n_s(1)-n_i(s,1))}\Big]\nonumber \\
&=&\prod_{i \in V\setminus S}\prod_{s\in
S(i)}\Big[A(s,i)^{n_i(s,1)}\times \nonumber
\\ &&(1-A(s,i)\beta_i)^{(n_s(1)-n_i(s,1))} \Big]. \nonumber
\end{eqnarray}

\noindent The log-likelihood function takes the following form:

\begin{eqnarray}
&&L(P(A(s,i))=\sum_{i \in V\setminus S}\sum_{s \in S(i)} \Big[ n_i(s,1)\log A(s,i) \nonumber \\
&&+(n_s(1)-n_i(s,1))\log(1-A(s,i)\beta_i)\Big ] \label{pathlikely}.
\nonumber
\end{eqnarray}

\noindent Differentiating the log-likelihood function {\it wrt}
$A(s,i)$ and letting the derivative be 0, we have

\begin{eqnarray}
A(s,i)=\dfrac{\gamma_i(s)}{\beta_i}, \mbox{\hspace{1cm}} s \in S(i).
\nonumber
\end{eqnarray}
\end{IEEEproof}

(\ref{pathsetpoly}) is almost identical to that obtained for the tree
topology except that the $\beta_i$ is consistent to all $\gamma_i(s), s \in
S(i)$. We call this the consistent condition since it ensures the
estimates obtained for $A(s, j), j \in a^s(i), s \in S(i)$ consistent
with each other. The condition is also implicitly presented in
(\ref{mgeneral}) that indicates there is an order in the estimation of a
general network, where $\beta_i, i \in J$ should be estimated before
 $A(s, k)$, $k \in a^s(i), s \in S(i)$.

 To have the MLE $\hat A(s,i), s \in S(i), i \in J$ satisfying the consistent condition, we have the following two theorems.
\begin{theorem} \label{jointnodetheorem}
Let $A(k,i)$ be the pass rate of the path connecting $k, k \in S(i)$
to $i, i \in J$ in a network of the general topology. There is a
polynomial, $H(A(k, i), S(i))$, as follows to express the estimate of
$A(k,i)$.
\begin{eqnarray}
&&H(A(k, i), S(i)) = 1-\dfrac{\hat\gamma_i(k)}{A(k,i)} - \nonumber \\
&& \prod_{j \in d_i}(1-\dfrac{\hat\gamma_i(k)\sum_{s \in S(i)} n_j(s,
1)}{A(k, i)\cdot \sum_{s \in S(i)} n_i(s,1)})=0 \label{mgeneralpoly}
\end{eqnarray}
where

\begin{equation}
\hat\gamma_i(k)=\dfrac{n_i(k, 1)}{n^k} \nonumber
\end{equation}

\end{theorem}

\begin{IEEEproof}
Assume  source $k, k \in S(i)$ sending probes to node $i$. Based on
the first equation of (\ref{mgeneral}), we have

\begin{eqnarray}
A(k,i)\beta_i = \dfrac{n_i(k,1)}{n^k}=\hat\gamma_i(k),
\mbox{\hspace{0.5 cm} }k \in S(i) \nonumber  \label{2connect}
\end{eqnarray}

\noindent  Since $|S(i)|>1$, there are more than one sources sending
probes to node $i$, the pass rates of two sources, $s$ and $k, s, k
\in S(i)$, to node $i$ are correlated that can be expressed as
\begin{eqnarray}
A(s,i)=A(k,i)\dfrac{\hat\gamma_i(s)}{\hat\gamma_i(k)},
\mbox{\hspace{0.5cm} } s, k \in S(i)  \label{2relation}
\end{eqnarray}

\noindent  Let $n_i^*(1)$ be the total number of probes reaching node
$i$, we have

\begin{eqnarray}
n_i^*(1) = \dfrac{A(k, i)}{\hat\gamma_i(k)}\sum_{s \in S(i)} n^s
\hat\gamma_i(s) \nonumber \label{totalnum}
\end{eqnarray}

\noindent Then, we have two ways to express $1-\beta_i$. If the
observations of the subtrees rooted at node $i$ satisfy
(\ref{difference}), we have
\begin{eqnarray}
1-\beta_i=\prod_{j \in d_i} \big[1-\dfrac{\sum_{s \in S(i)} n_j(s,
1)}{n_i^*(1)}\big] \label{subloss}
 \end{eqnarray}

\noindent and using (\ref{pathsetpoly}), we have

 \begin{eqnarray}
  1-\beta_i
  &=& 1- \dfrac{\hat\gamma_i(k)}{A(k,i)} \nonumber
  \end{eqnarray}
  \noindent Connecting the two, we have
\begin{eqnarray}
1- \dfrac{\hat\gamma_i(k)}{A(k,i)}  &=&\prod_{j\in d_i}
  \Big(1-\dfrac{\hat\gamma_i(k) \cdot \sum_{s \in S(i)} n_j(s, 1)}{A(k,i)\cdot \sum_{s \in S(i)}
  n^s
\hat\gamma_i(s)}\Big) \nonumber  \label{pathrateeql}
  \end{eqnarray}

\noindent Except for $A(k, i)$, all others, e.g. $\hat\gamma_i(k),
n_j(s,1)$, are either known or obtainable from observations. Thus, the
above equation is a polynomial of $A(k, i)$. Alternatively, using
(\ref{subloss}) to replace $\beta_i$ from (\ref{mgeneral}), we have
the same result.
\end{IEEEproof}
(\ref{mgeneralpoly}) generalizes (\ref{minc}) that considers various
paths ending at a common node, including those having $|S(i)|\geq1$.
For node $i$ having $|S(i)|= 1$, (\ref{mgeneralpoly}) degrades to
(\ref{minc}). For $|S(i)|>1$, if $j\in d_i$,
\begin{equation}
\dfrac{\sum_{s \in S(i)} n_j(s, 1)}{\sum_{s \in S(i)} n_i(s,1)}
\label{alphai}
\end{equation}
is the estimate of the pass rate of the link connecting node $i$ to
node $j$, where the numerator is the sum of the probes sent by $S(i)$
that reach $Rs(i)$ and the denominator is the sum of the probes sent
by $S(i)$ that reach $Rs(j), j \in d(i)$. Both are obtainable from
$\Omega$. The estimate is built on the arithmetic mean that considers
the contribution of all sources sending probes to the link.

 Solving (\ref{mgeneralpoly}), we have $\widehat A(k, i)$.
Then, we can have $\widehat A(s, i), s \in S(i)\setminus k$ from
(\ref{2relation}) and
 \[
\hat\beta_i=\dfrac{n^s}{\widehat A(s,i)}, s \in S.
\]
Using Lemma 1 in \cite{CDHT99}, we are able to
prove there is only one solution to (\ref{mgeneralpoly}) in $(0,
1)^m$,
 To prove the estimate obtained from (\ref{mgeneralpoly}) is the MLE, we
resort on a well known theorem for the MLE of a likelihood function
yielding the exponential family.
\begin{theorem}\label{properties of statistics1}
If a likelihood function  belongs to a standard exponential family
with $A(s,i)$ as the natural parameters, we have the following
results:
\begin{enumerate}
  \item the likelihood equation $\frac{\partial
L(\theta)}{\partial \theta_i}=0$ has at most one solution
$\theta_i^*\in \Theta$;
  \item if $\theta_i^*$ exists, $\theta_i^*$ is the MLE.
\end{enumerate}
\end{theorem}
\begin{IEEEproof}
This theorem can be found from a classic book focusing on exponential
families, such as \cite{Brown86}. The likelihood function presented in
(\ref{pathlikely}) belongs to the exponential family, where $A(s,i)$
is the natural parameters. Then, the estimate obtained from
(\ref{mgeneralpoly}) is unique in its support space and it is the MLE
of $A(s,i)$.
\end{IEEEproof}

\section{Solutions}
\label{section6}
Theorem \ref{jointnodetheorem} is not only valid to estimate the pass
rates of a path connecting a source to a joint node, but also valid to
estimate the pass rate of a path connecting a source to a node in an
intersection. Then, the loss rate of a link in an intersection can be
obtained easily since (\ref{theta2A}) is the bijection function
$\Gamma$ from $\Theta$ to $A$, and then $\Gamma^{(-1)}$, from $A$ to
$\Theta$, is as follows

\begin{equation}
1-\theta_i=\dfrac{\sum_{s \in S(i)} n^s A(s, i)}{\sum_{s \in S(i)} n^s
A(s, f(i))}. \end{equation} Unfortunately,  $\Gamma^{(-1)}$ cannot be
applied to the links falling into SBRL and having at least a
descendant connecting an intersection, i.e. the ancestors of an
intersection, because of the consistent condition.

\subsection{Consistency and Decomposition}

To estimate the loss rate of link $j, j \in a(i) \bigwedge i \in J$,
called ancestor links, a divide-and-conquer strategy is proposed here
that is not only an algorithm to estimate the loss rates of a general
network, but also a necessary measure to ensure the consistency
established in theorem \ref{general MLE}. Instead of directly finding
a likelihood equation for the ancestor links from the statistics
proposed in Section \ref{general sufficient stat}, a general network
covered by a number of trees is divided into a number of independent
ones before being estimated. The strategy is built on equation
(\ref{mgeneralpoly}) that shows the parameter estimation in a
hierarchical structure with probes flowing in one direction can be
divided into a number of sections as the d-separation stated in
\cite{JP89}.

Given the number of probes reaching a node, the descendants of the
node not only become independent from each other, but also become
independent from their ancestors. Since each of the independent parts
is a tree, the estimators proposed for the tree topology can be used
to estimate the loss rates of the independent parts. Then, the
immediate question is which nodes should be selected as the
decomposing points that can minimize the cost of estimation. Two
criteria are proposed here for the optimal decomposition, which
should:
\begin{enumerate}
\item  ensure that all the nodes and links in an independent tree receive probes from the same set of
sources. Then the sources can be considered a virtual source sending
probes to the receivers of the independent tree; and
\item minimize the number of independent trees created after decomposition since the computation cost is proportional to
the number of the independent trees.
\end{enumerate} The following theorem is presented for the optimal strategy to
select decomposing points.

\begin{theorem}
Given a general network covered by multiple trees, the optimal
strategy to decompose the overlapped multiple trees into a number of
independent trees is to use the nodes of $J$ as decomposing points.
\end{theorem}
\begin{IEEEproof}
Obviously the nodes or  links of an intersection can only receive
probes from the same set of sources. The minimum number of independent
trees is proved by contradiction. Firstly, we assume there are two
strategies, say A and B. Strategy A decomposes a general network into
a number of independent subtrees and there is at least one of the
decomposing points that is not a joint point and the total number of
the independent subtrees obtained is $m'$. On the other hand, strategy
B decomposes the network at the joint points only and the total number
of independent subtrees is $m$. If strategy A is better than strategy
B, we should have $m' < m$. However, this is impossible: given the
fact that all of the subtrees rooted at a joint point can only become
independent from each other if we know the state of the joint point
according to d-separation. Then, we have $m' \geq m+1$, which
contradicts to the assumption. Then, the theorem follows.
\end{IEEEproof}

To start the the divide-and-conquer strategy, we need to have a method
to estimate $n_i^*(1), i \in J$. If $i$ has more than 6 descendants,
there is not a closed form solution according to Galois theory. Then,
an approximation method, such as (\ref{mgeneralpoly}) or the EM
algorithm, is needed to find the solution. Apart from them, an
iterative procedure based on fixed point theory can also accomplish
this. The iterative procedure uses:

\begin{eqnarray}
&&(1-\beta_i)^{(q+1)}= \nonumber \\ &&\prod_{j \in d_i}
\Big[\frac{\sum_{k \in S(j)}n^k_{j}(0)}{\sum_{k \in S(i)}
n^k_i(1)}+\frac{\sum_{k \in S(j)} n^k_j(1)}{\sum_{k \in S(i)}
n^k_i(1)}\cdot(1-\beta_i)^{(q)}\Big]
\nonumber \\
 \label{munique}
\end{eqnarray}

\noindent to compute $(1-\beta_i)^*$ \cite{todd76}.  According to Lemma 1 in \cite{CDHT99}, there is only one $(1-\beta_i)$ in $(0, 1)$
satisfying (\ref{mgeneralpoly}). Then, the iterative procedure is
assured to coverage at $(1-\beta_i)^*$  and the number of iterations
depends on the initial value of $(1-\beta_i)$, $(1-\beta_i)^{(0)}$. If
the initial value is selected properly, i.e. close to the fixed point,
only a few of iterations are needed. A number of methods can be used
to obtain the initial value, one of them is to use a weighted average
obtained from the estimates of the trees involving in subtree $i$;
another is to use the top down algorithm proposed in \cite{ZG05}. In
our experience for a network as Figure \ref{multisource}, the
procedure needs 5.25 iterations on average to converge to the fixed
point when $\epsilon$ is set to $10^{-10}$. Alternatively, Newton
Raphson algorithm can be applied to (\ref{munique}) to find the
$(1-\beta_i)^*$ that can converge to $(1-\beta_i)^*$ quickly as well.
Given $(1-\beta_i)^*$, we have $\beta_i^*=1-(1-\beta_i)^*$ and
$n_i^*=\dfrac{n_i(1)}{\beta_i^*}$.

As stated, the independent trees obtained from decomposition can be
divided into two groups on the basis
 of ingress or egress of the links connecting or connected to the decomposing points. The group consisting of the trees that have at least an ingress
 link to a joint node is called
 the ancestor group, all others that have an egress link from a joint node is called the descendant
 group.
For the descendant group, given $n^*_i(1)$,  the maximum likelihood
estimators developed for the tree topology can be applied to estimate
the loss rates of the independent trees.

Although the trees in the  ancestor group as those in the descendant
group are independent from each other given $n^*_i(1)$, they cannot be
estimated directly by an estimator developed for the tree topology. As
previously stated, the pass rate of a path ending at a node that is an
ancestor of a joint node depends on the estimate obtained for the
intersection. Then, the pass rate of the path connecting source $r, r
\in S(i)$ to node $i, i \in J$ should be $\dfrac{\gamma_i(r)}{\hat
\beta_i}$. If $j$ is the last link of the path, i.e  the link ending
at node $i$, the statistic of link $j$, $n_j^r(1)$, is equal to
\begin{equation}
\hat n_j^r(1)=\dfrac{n^r_j(1)}{\beta_i^*}, j \in p(i). \label{estimate
ni}
\end{equation}

\noindent  where $p(i)$ denotes the parents of node $i$. To ensure the
consistency of estimation with the pass rates from $k$ to $a^k(l), l
\in p(i), k \in S(i)$, we need to use $\hat n_j^r(1)$ instead of
$\Omega_r(j)$. Then, (\ref{polyequal}) is no longer valid to the
estimation of the links of the ancestor group and we need to use the
first equation of (\ref{general}) in the subsequent estimation. As
previous stated, we need to consider how to express $\beta_q, q \in
a^r(j)$ to have a valid likelihood equation according to $\Omega_q(s),
s \in S(i)$. Under the prefect assumption of $\Omega_q(s), s \in S(i)$
\cite{Zhu11}, we use $\hat A(s,i)=\dfrac{\hat n_j^s(1)}{n^s}$ to
replace $\gamma_i(s)$ in the modified path-based estimator that has
the following form:

\begin{eqnarray}
1-\dfrac{\gamma_{f^s(i)}(s)}{A(s, f^s(i))}&= &\big[1-\dfrac{\hat A(s,
i)}{A(s, f^s(i))}\big] \nonumber \\
&&\prod_{j \in d_{f^s(i)} \setminus
i}(1-\dfrac{\gamma_j(s)}{A(s,f^s(i))} ). \label{rateestimator}
\end{eqnarray}

\noindent Note that (\ref{rateestimator}) is obtained under the basis
of (\ref{difference}), that is a polynomial of $A(s, f^s(i))$. Once
having $A(s, f^s(i))$, we are able to estimate $n^*_{f^s(i)}(1)$, then
we move a level up toward $s$ and use (\ref{rateestimator}) to
estimate $A(s, f^s_2(i))$. This process continues from bottom up until
reaching $s$. If
 the tree being estimated has more than one intersections, at
the common ancestors of the intersections, the RHS of
(\ref{rateestimator}) will have a number of the left-most terms, one
for an intersection plus the product term for those subtrees that do
not intersect with others. If the total number of intersections plus
the number of independent subtrees is larger than 5, there is no a
closed form solution to (\ref{rateestimator}).

Compared to the tree topology, the general topology, despite having a
similar likelihood equation, has its unique features in estimation and
one cannot simply use the estimators proposed for the tree topology in
the estimation of the general topology.

\section{Statistical Property of the Estimators}
\label{section7}
Apart from providing solution to the general topology, we study the
statistical properties of the solution, such as whether the estimators
are minimum-variance unbiased estimators (MVUE); and/or the estimate
obtained is the best asymptotically normal estimates (BANE), etc. This
section provides the results obtained from the study.
\subsection{ Minimum-Variance Unbiased Estimator}
The estimators proposed in this paper are MVUE and the following
theorem proves this.
\begin{theorem} \label{MVUE theorem}
The estimator proposed in this paper is MVUE and the variances of the
estimates reach the Carm\'{e}r-Rao bound.
\end{theorem}
\begin{IEEEproof}
The proof is based on Rao-Blackwell Theorem that states that if g(X)
is any kind of estimator of a parameter $\theta$, then the conditional
expectation of g(X) given T(X), where T is a sufficient statistics, is
typically a better estimator of $\theta$, and is never worse. Further,
if the estimator is the only unbiased estimator, then, the estimator
is the MVUE.

To prove the estimator is an unbiased estimator, we use (\ref{general}) instead of (\ref{minc}) since  the  former is not only more
general than the latter, but also much simpler than the latter in the
proof. (\ref{general}) can be written as
\begin{equation}
1-\hat\theta_i = \dfrac{\dfrac{ n_i(1)}{\hat n_{f(i)}(1)}}{\hat
\beta_i}
\end{equation}
Let $\bar{\theta_i}=1-\theta_i$. We then have
\begin{eqnarray}
&&E\Big ( \dfrac{\dfrac{ n_i(1)}{\hat n_{f(i)}(1)}}{\hat\beta_i}-\bar{\theta_i} \Big ) \nonumber \\
&=& E\Big (\dfrac{\dfrac{ n_i(1)}{\hat n_{f(i)}(1)}}{\hat \beta_i}\Big ) - E(\bar{\theta_i}) \nonumber \\
&=& E\Big ( \dfrac{ n_i(1)}{\hat n_{f(i)}(1)} \cdot \dfrac{
\hat n_i(1)}{n_i(1)}\Big )- E(\bar{\theta_i}) \nonumber \\
 &=& E\Big ( \dfrac{\hat n_i(1)}{\hat n_{f(i)}(1)} \Big )-
 E(\bar{\theta_i}) \nonumber \\
 &=& E\Big(\dfrac{1}{\hat n_{f(i)}(1)}\sum_{j=1}^{\hat n_{f(i)}(1)}x_j\Big)-
 E(\bar{\theta_i}) \nonumber \\
 &=& \theta_i -\theta_i =0
\end{eqnarray}
The statistics used in this paper has been proved to be the minimal
complete sufficient statistics.  Then, applying Rao-Blackwell theorem,
the theorem follows.

Given theorem \ref{MVUE theorem}, it is easy to prove the variance of
the estimates, e.g. $\hat\theta_i$ and $\hat \beta_i$, obtained by
(\ref{mgeneralpoly}) are equal to Carm\'{e}r-Rao low bound since
(\ref{pathlikely}), the likelihood function, belongs to the standard
exponential family.
\end{IEEEproof}
Based on Fisher information we can prove the variance of the estimates
obtained by (\ref{mgeneralpoly}) from $\Omega$ is also smaller than
that of an estimate obtained by (\ref{minc}) from $\Omega_s, s \in S$.
Since the receivers attached to intersections observe the probes sent
by multiple sources, the sum of the observed probes is at least larger
than or equal to the maximum number of probes observed from a single
source. Therefore, there is more information about the loss rates of
the links located in the intersections since information is addictive
under {\it i.i.d.} assumption. With more information, the variance of
an estimate of a parameter must be smaller than another obtained from
the probes sent by a single source according to Fisher information.
Given the less varied estimates from the intersections, the variances
of the estimates of other links that are not in an intersection are
also reduced, at least not increased. Therefore, the estimates
obtained by (\ref{mgeneralpoly}) is better than those obtained from a
single source, which also implies the necessity of an order in the
estimation of a general network.

\subsection{General Topology vs. Tree Topology}

The results presented in this paper can be viewed as a generalization
of the results presented in \cite{CDHT99} since the tree topology is
only a special case of the general topology. The findings and
discovery presented in this paper cover those presented in
\cite{CDHT99}. The following corollary confirms this:
\begin{corollary}\label{special case}
Any discovery, including theorems and algorithms, for loss estimation
in the general topology, holds for the tree topology as well.
\end{corollary}

For instance, (\ref{mgeneralpoly}) obtained for the general topology
holds for the tree one. When $S(i)=\{k\}$,  we have
\begin{eqnarray}
&&H(A(s, k), S(k)) = 1-\dfrac{\gamma_i(k)}{A(k,i)} - \nonumber \\ &&
\prod_{j \in d_i}(1-\dfrac{\gamma_i(k)\sum_{s \in S(i)} n_j(s,
1)}{A(k, i)\cdot \sum_{s \in S(i)} n_i(s,1)}) \nonumber \\
&=&1-\dfrac{\gamma_i(k)}{A(k,i)} -\prod_{j \in
d_k}(1-\dfrac{\gamma_i(k) n_j(1)}{A(k,i)\cdot n_i(1)}) \nonumber \\
&=&1-\dfrac{\gamma_i(k)}{A(k,i)} -\prod_{j \in
d_k}(1-\dfrac{\gamma_j(k)}{A(k,i)})=0
\end{eqnarray}

\noindent the last equation is  $H(A_i, i)$ presented in
\cite{CDHT99}.

 On the other hand, we are also interested in whether
those properties discovered from the tree topology can be extended
into the general one and the difference between the original
properties and the extended ones, in particular for the rates of
convergence.

\subsection{Large Sample Behavior of the Estimator}

Since the estimate obtained by the proposed estimator is the MLE $\hat
\theta_i$, we are able to apply some general results on the asymptotic
properties of MLEs in order to show that $\sqrt n(\hat \theta_i -
\theta_i)$ is asymptotically normally distributed as $n\rightarrow
\infty$. Using this, we can estimate the number of probes required to
have an estimate with a given accuracy for many applications. The
fundamental object controlling convergence rates of the MLE is the
Fisher Information Matrix at $\theta_i$. Since $L(\theta)$ yields
exponential family, it is straightforward to verify that $\hat
\theta_i$ is consistent and that $L(\theta)$ satisfies conditions
under which $\textit{I}$ is equal to

\[
I_{jk}(\theta)=-E\dfrac{\partial^2L}{\partial \theta_j \partial
\theta_k}(\theta)
\]

\noindent Eliminate singular on the boundary of $(0,1]^{|E|}$, we have
\begin{theorem}\label{asympototic theorem}
When $\theta_i \in (0,1), i \in V\setminus S, \sqrt n(\hat \theta_i -
\theta_i)$ converges in distribution as $n\rightarrow \infty$ to an
$|E|$ dimensional Gaussian random variable with mean 0 and covariance
matrix $I^{-1}(\theta)$, i.e.
\[
\sqrt n (\hat\theta -\theta) \xrightarrow{D} N(0,I^{-1}(\theta))
\]

\noindent and $\hat \theta_i$ is the best asymptotically normal
estimate (BANE).
\end{theorem}

\begin{IEEEproof}
\noindent It is know that under following regularity conditions:
\begin{itemize}
\item the first and second derivatives of the log-likelihood function
must be defined.
\item the Fisher information matrix must not be zero, and must be
continuous as a function of the parameter.
\item the maximum likelihood estimator is consistent.
\end{itemize}
\noindent the MLE has the characteristics of asymptotically optimal,
i.e., asymptotically unbiased, asymptotically efficient, and
asymptotically normal. The characteristics are also called BANE.

It is clear that (\ref{pathlikely}), the likelihood function used in
this paper, belongs to the standard exponential family, which ensures
the consistence and uniqueness of the MLE. To satisfy  the second
condition for the exponential family, $n_i(s, 1), i \in E, s \in S$
should not be linearly related, this is true as $n_i(s,1), i \in E, s
\in S$ have been proved to be the minimal sufficient statistics. Then,
we only need to deal with the first condition. Obviously,
(\ref{mgeneral}) has both first and second derivatives in
$(0,1)^{|E|}$, and $L(\theta)$ is strictly concave, which ensures the
Fisher information matrix $I(\theta)$ positive definiteness.
\end{IEEEproof}

Theorem  \ref{asympototic theorem} states such a fact that with the
increase of the number of probes sent from sources, there are more
probes reaching the links of interest. Then, there is more information
for the paths to be estimated. Let $I_0(\theta)$ is the Fisher
Information for a single observation, we have
$I(\theta)=nI_0(\theta)$, where $n$ is the number of observations
related to the link/path being estimated.

As $n\rightarrow \infty$, the difference in terms of Fisher
information between the two estimators approaches to zero. Therefore,
as $n \rightarrow \infty$, estimation can be carried out on the basis
of individual tree and the asymptotical properties obtained previously
for the tree topology \cite{CDHT99} hold for the general topology as
well. On the other hand, if $n<\infty$, the estimate obtained by
(\ref{mgeneral}) or (\ref{mgeneralpoly}) is more accurate than those
obtained from an individual tree. The simulation results presented in
the next section illustrate this that shows the fast convergence of
the estimates for the links located in the intersection because there
are more information about the links.

 Although there are a number of large sample properties that can
be related to loss inference in the general topology, including
various asymptotic properties, we would not discuss them further
because $n \rightarrow \infty$ means a large number of probes be sent
from sources to receivers that requires a long period of stationarity
of the network, including traffic and connectivity, which is
impractical based on the measurement \cite{PAX99}.

\section{Simulation Study}
\label{section8}

For the purpose of proof of concept, a series of simulations were
conducted on a simulation environment built on {\it ns2}, the network
simulator 2 \cite{NS2}. The network topology used in the simulations
is
 shown in Figure \ref{multisource}, where two sources located at node
0 and node 16
 multicast probes  to the
 receivers attached to the leaf nodes. Both use a constant rate to send probes to the receivers, where the interval between
 2 probes is set to 0.01 second. The binary structure
used here is for the simplicity reasons. Apart from the traffic
created by probing, a number of TCP flows with various window sizes
and a number of on/off UDP flows with various burst rates and on/off
periods are added at the roots and internal nodes that acting as
background traffic send packets to the receivers attached to the leaf
nodes. Each simulation run for 200 times and each time a random seed
is selected to start the simulation. The samples collected in a run
vary from 200 to 2000, with an interval of 200, to measure the impact
of the sample sizes on the accuracy of estimation. The accuracy of
estimation is measured by the relative error that is defined as:

\[
\dfrac{abs(\mbox{actual loss rate - estimated loss
rate})}{\mbox{actual loss rate}}.
\]
To compare the estimates obtained from end-to-end observation with
those obtained directly at each node, an agent is added at each node
to record the number of packets lost and passed a link. The ratio
between the losses of a link and the sum of the losses and passes of
the link is used as the actual loss rate of the link. As stated, the
impact of background traffic on the accuracy of estimation is also a
concern of this study. The next two subsections are devoted to these
two issues.

\subsection{Impact of Sample Size on Accuracy}

In the first round simulation, there are 9 TCP flows in the multicast
tree rooted at node 0, where

\begin{itemize}
\item 4 from node 0 to nodes 8 and 9;
\item 2 from node 0 to nodes 10 and 11; and
\item 3 from node 2 to nodes 9, 10 and 11.
\end{itemize}
In addition, there 12 UDP flows in the multicast tree, including

\begin{itemize}
\item 2 flows from node 0 to
node 8 and node 9;
\item 4 flows from node 3 to nodes 12-15;
\item 2 flows from node 2 to nodes 10 and 11;
\item 2 flows from node 1 to nodes 10 and 11; and
\item 2 flows from node 4 to nodes 8 and 9, respectively.
\end{itemize}
 The window size used by the TCP flows is 50.

 For the multicast tree rooted at node 16, there are 6 TCP flows
and 4 UDP ones. Among the 6 TCP flows, 2 of them are from node 17 to
nodes 8 and 9, the other 4 are from node 16 to nodes 21-24,
respectively. The UDP flows are from node 18 to nodes 21-24. The
window size of the flows is set to 60. The loss rate of each link is
controlled by a random process that has $1\%$ drop rate.

\begin{figure}
\begin{center}
\epsfig{figure=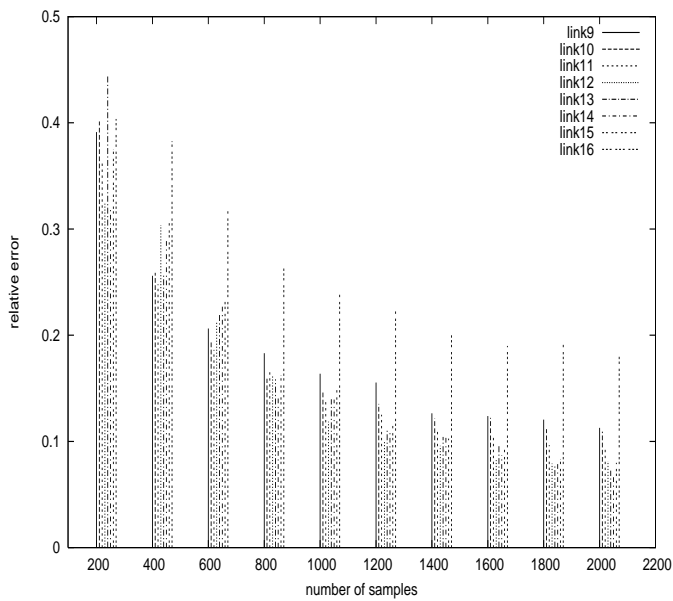,height=9.0cm,width=8.0cm,angle=-90}
 \caption{Relative Error for Link 9-16}
\label{rel9-16}
\end{center}
\end{figure}

The resultant relative errors of the 24 links are presented in Figure
\ref{rel1-8} to Figure \ref{rel17-24},  8 links in each. The figures
show that with the increase of samples, the relative errors are
decreased as expected. The relative errors of  links 0 to 15 drop to
around 10\% when 2000 samples are used in estimation. This phenomenon
is consistent with the expectation, i.e. the number of samples used in
estimation is inversely proportional to the variance of an estimate.
However, whether we are able to use more samples to achieve the
required accuracy is a question that has not been answered previously.
What we found from the simulation is that the loss process of a link
is not independent from background traffic. Then, we need to
investigate traffic dependent models in the future.

\subsection{Impact of Background Traffic on Accuracy}

\begin{figure}
\begin{center}
\epsfig{figure=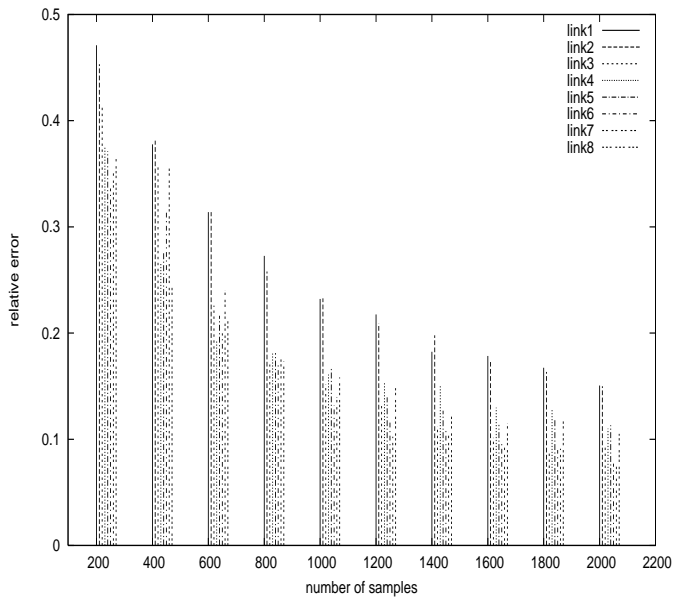,height=9.0cm,width=8.0cm,angle=-90}
\caption{Relative Error for Link 1-8} \label{rel1-8}
\end{center}
\end{figure}

\begin{figure}
\begin{center}
\epsfig{figure=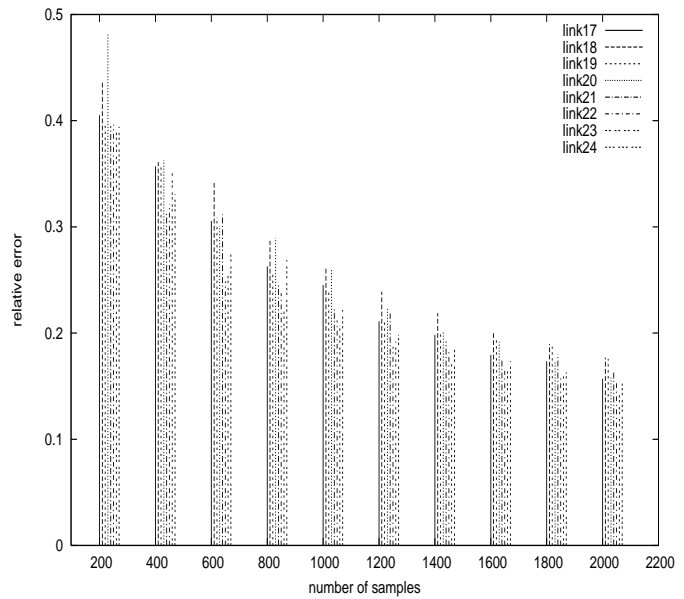,height=9.0cm,width=8.0cm,angle=-90}
 \caption{Relative Error for Link 17-24}
\label{rel17-24}
\end{center}
\end{figure}

Comparing Figure \ref{rel1-8} with Figure \ref{rel17-24}, one is able
to notice the difference between them, where the relative error of
Figure \ref{rel17-24} is significantly larger than that of Figure
\ref{rel1-8}. Another interesting point between the two figures is the
improvement rate of the accuracy against the number of samples, where
the rate appears gradually reduced as the sample sizes in Figure
\ref{rel17-24}. When the sample size reaches 1600, further increasing
the number of samples only has a marginal impact on the accuracy. This
phenomenon triggers us to consider the causes of the reduction since
the network structure used in the simulation is symmetric and the
parameters used for each link is identical. Apart from those,  both
sources use the same rate to send probes to the receivers. The only
difference between them is the background traffics flowing on the
links.  The traffic flowing on the subtrees rooted at node 3 consists
of a number of on/off UDP flows. In contrast, there are four FTP flows
sending packets from node 16 to nodes 21, 22, 23, and 24,
respectively, plus four on/off UDP flows forwarding traffic to the 4
receivers from node 18. Thus, there are two possible causes that lead
to the difference:

\begin{itemize}
\item a mismatch between the traffic patterns of the probe flow and the
background flows, and \item a mismatch between the traffic intensities
of the two type of flows. \end{itemize}
 It is hard to overcome the
former if the loss model of a link is related to traffic unless
knowing the correspondence between models and traffic in advance and
having an estimator for each model. Without them, increasing the
number of samples is the only option that may improve the accuracy in
some degree, but it also has a side-effect as previous stated. To have
more probes for estimation requires the traffic remains stable. If the
duration of stability is also an issue, what we can do is to increase
the intensity of probing, i.e. sending more probes in a period.
However, the probing may intervene the background traffic that can
result in an incorrect estimate.

\begin{figure}
\begin{center}
\epsfig{figure=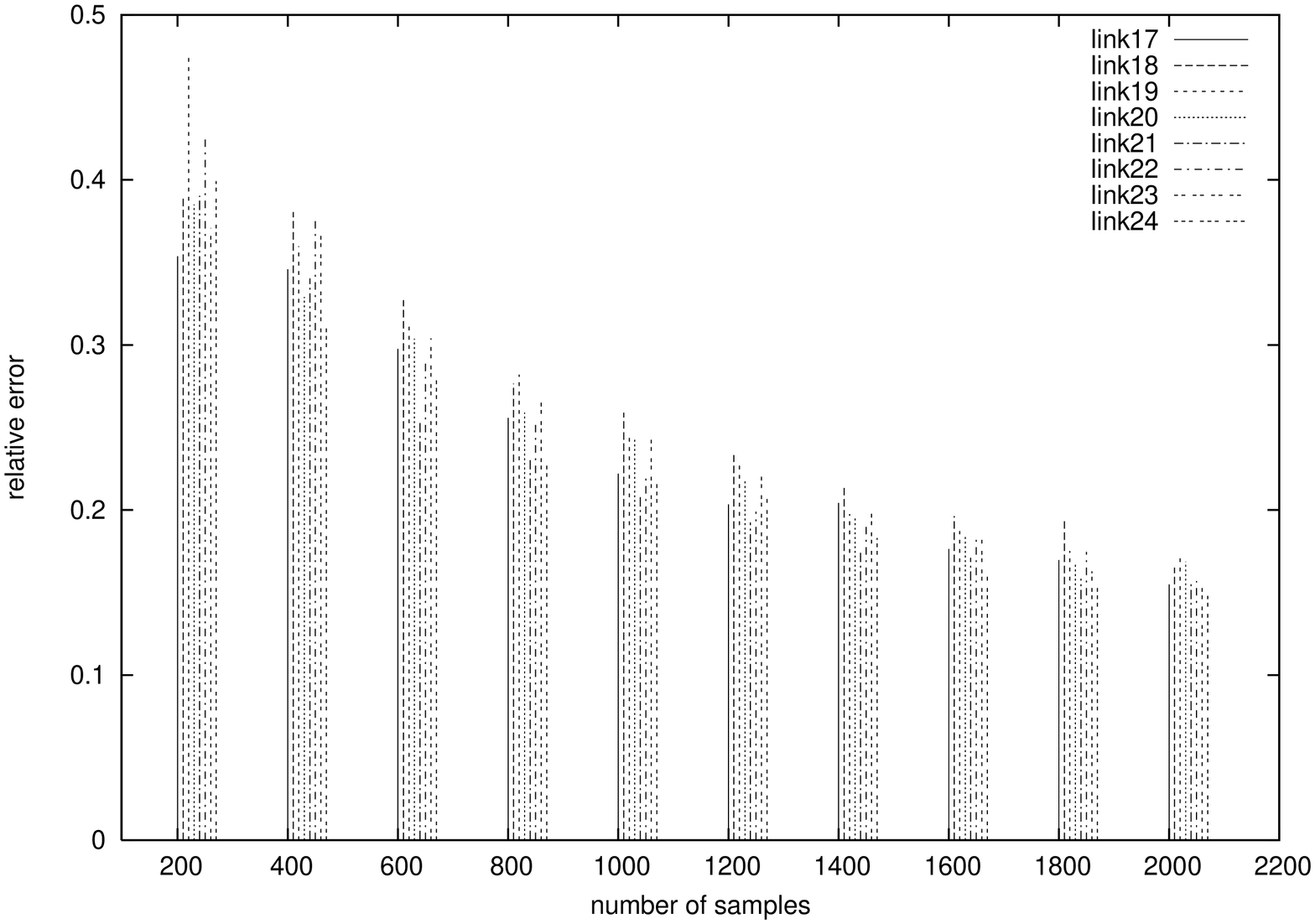,height=9.0cm,width=8.0cm,angle=-90}
 \caption{Relative Error of Link 17-24 after removing 4 ftp flows from node 16}
\label{rel17-24withlowflow}
\end{center}
\end{figure}

\begin{figure}
\begin{center}
\epsfig{figure=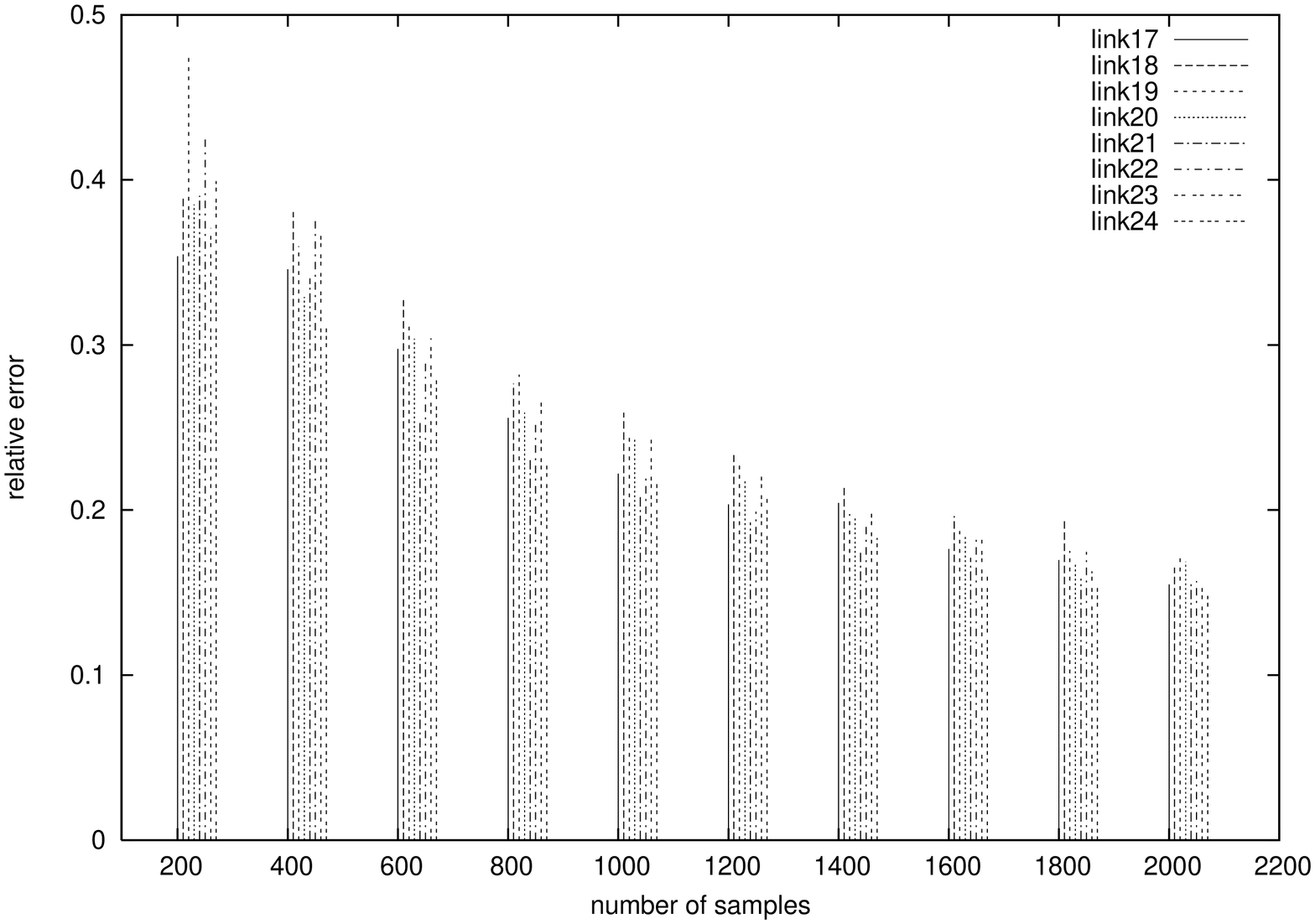,height=9.0cm,width=8.0cm,angle=-90}
 \caption{Relative Error of Link 17-24 with 2 TCP flows}
\label{rel17-24withmiddleflow}
\end{center}
\end{figure}

\begin{figure}
\begin{center}
\epsfig{figure=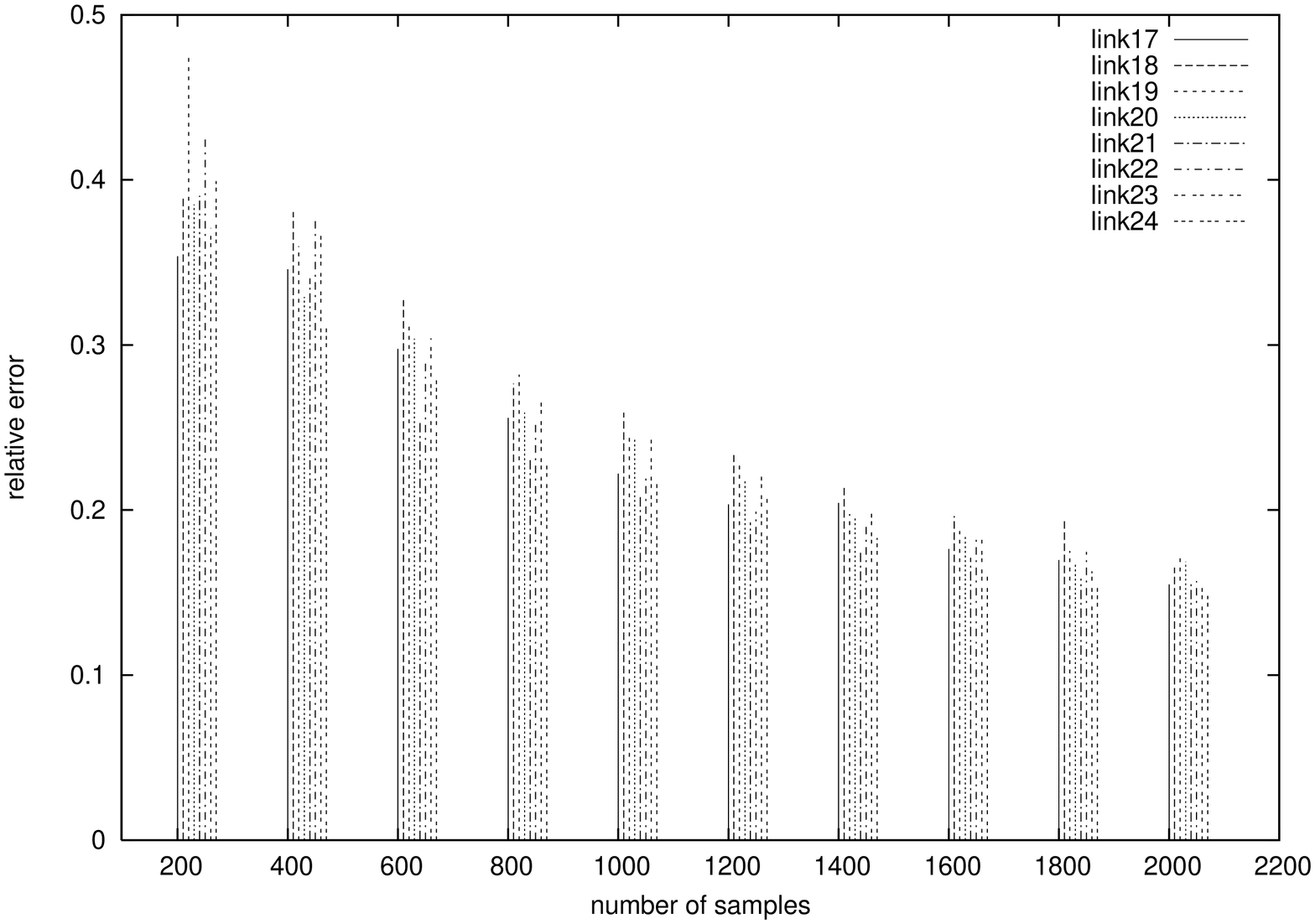,height=9.0cm,width=8.0cm,angle=-90}
 \caption{Relative Error of Link 17-24 with 2 half-window TCP flows}
\label{rel17-24withhalfwin}
\end{center}
\end{figure}

To confirm the above, two rounds of simulations are carried out: both
stem from the previous setting, one removes the 4 TCP flows from node
16 and the other removes two of the 4 TCP flows. The result of the
former is presented in Figure \ref{rel17-24withlowflow} that clearly
shows the improvement in terms of accuracy. The relative error remains
steady at around 5\% for links 18-24, irrespective to sample sizes.
This is because of the consistence between the models used by the
estimator and the background traffic. In contrast, when  2 of the 4
TCP flows are restored, one sends probes from node 16 to node 21 and
the other to node 23,  the relative errors of the links having TCP
traffic are bounced back to where they were, as shown in  Figure
\ref{rel17-24withmiddleflow}. However, the relative errors of link 22
and link 24 remain at the same level as those in Figure
\ref{rel17-24withlowflow}. Further, we halve the window size of the
TCP flows to see its impact on the accuracy. The result is presented
in Figure \ref{rel17-24withhalfwin}. Comparing Figure
\ref{rel17-24withhalfwin} with Figure \ref{rel17-24withmiddleflow},
there is little difference between them. This reflects the loss
process created by TCP flows are similar to each other that is
independent from the number of flows and size of congestion window.

 The results presented
in this subsection indicate that apart from the sample size, to have
an accurate estimate we must consider the correspondence between the
model used in estimation and that of traffic. Without the knowledge,
we can only adjust the ratio between the probes sent by a source and
that of the ongoing packets to improve estimation accuracy.  For
instance, in the first round, the background traffic for the shared
subtree is much higher than other links and with both TCP and UDP.
Nevertheless, the relative errors of the links in the intersection can
reach about 10\% when 4000 samples are used in estimation that are
significantly better than those of links 18-24 since the sending rate
of probes for those links is two times of others.



\section{Conclusion}

In this paper, the findings obtained recently on loss tomography are
presented that include theoretical findings and practical algorithms.
In theory, the internal view introduced in the paper leads to a set of
complete minimal sufficient statistics that contain all the
information needed to compute the loss rates of a network. Based on
the statistics, the frequently used likelihood function is rewritten,
and subsequently a set of likelihood equations is obtained. Solving
the likelihood equations, a direct expression of the MLE is obtained
for the link-level loss rates of the tree topology. In contrast to the
previous works, the direct expression considers the dependency between
a likelihood equation and the data set obtained from experiment.
Because of the dependency, the direct expression proposed in this
paper is applicable to all data sets, while the previous ones is at
most applicable to one type of data set.

With the success achieved on the tree topology, two direct expressions
of the MLE are derived for the general topology, one is a link-based
estimator and the other is a path-based one. The former has a similar
structure as the one derived for the tree topology, which also ensures
most of the theorems obtained for the tree topology hold for the
general topology. The latter is a polynomial that has a similar
structure as (\ref{minc}). In fact, the latter generalizes the
path-based estimator proposed for the tree topology. In addition to
the likelihood equations, the dependency between estimates has been
identified in the estimation of the general topology that subsequently
leads to the introduction of estimation order for the general
topology. To impose the order in estimation, a divide-and-conquer
strategy is proposed that decomposes the trees used to cover a general
network into a number of independent trees. Apart from the
divide-and-conquer strategy, a number of methods are proposed to
estimate the number of probes reaching the roots of the independent
trees, including a fixed-point procedure. Further, the independent
trees obtained from decomposition are divided into two groups called
descendant and ancestor, respectively. The estimators developed for
the tree topology can be used on the descendant group, while a new
estimator is proposed to handle those falling into the ancestor group.

\end{document}